\documentclass[10pt,conference]{IEEEtran}
\IEEEoverridecommandlockouts

\usepackage{cite}
\usepackage{amsmath,amssymb,amsfonts}
\usepackage{algorithmic}
\usepackage{graphicx}
\usepackage{tabularx}
\usepackage{textcomp}
\usepackage{xcolor}
\usepackage{booktabs}
\usepackage{glossaries}
\usepackage{listings}

\usepackage{hyperref}
\usepackage{listings}
\usepackage{colortbl}
\usepackage{booktabs} 
\usepackage{enumitem}
\usepackage{caption}
\usepackage{amsmath}
\usepackage{balance}
\usepackage{tabularx}
\usepackage{minted} 
\usepackage{romannum}
\usepackage{textcomp}
\usepackage[T1]{fontenc}
\usepackage{newfloat}
\usepackage{tikz}
\usetikzlibrary{positioning}
\usetikzlibrary{tikzmark}
\usetikzlibrary{shapes.geometric}
\usetikzlibrary{calc}
\usetikzlibrary{
arrows.meta, 
backgrounds, 
fit, 
shadows
}
\usepackage{tabularx,booktabs}
\usepackage[most]{tcolorbox}
\usepackage[english]{babel}
\usepackage{url}

\usepackage{float}
\definecolor{shadecolor}{gray}{0.97}
\usepackage[linewidth=1pt]{mdframed} 
\usepackage{titlesec}
\usepackage{xcolor}
\usepackage{pgfplots}
\usetikzlibrary{calc}  
\usepackage{wrapfig}
\usepackage{subcaption}
\usepackage{fontawesome5}
\usepackage{svg}
\usepackage{microtype}

\definecolor{lightred}{RGB}{255,200,200}
\definecolor{codegreen}{rgb}{0,0.6,0}
\definecolor{codegray}{rgb}{0.5,0.5,0.5}
\definecolor{codepurple}{rgb}{0.58,0,0.82}
\definecolor{backcolour}{rgb}{0.95,0.95,0.92}

\newtcolorbox{takeawaybox}{
    colback=gray!8,
    colframe=black,
    boxrule=0.5pt,
    arc=2pt,
    left=4pt, right=4pt, top=3pt, bottom=3pt,
    title={\textbf{Key Takeaway}},
    fonttitle=\small,
    coltitle=black,
    attach boxed title to top left={yshift=-15pt, xshift=4pt},
    boxed title style={colback=white, colframe=black, boxrule=0.5pt, arc=1pt}
}

\makeglossaries

\newacronym{llm}{LLM}{Large Language Model}
\newacronym{zs}{ZS}{Zero Shot}
\newacronym{fs}{FS}{Few Shot}
\newacronym{cot}{CoT}{Chain of Thought}
\newacronym{tot}{ToT}{Tree of Thought}
\newacronym{sc}{SC}{Self Consistency}
\newacronym{gem04}{GEM04}{Gemma-3-4B}
\newacronym{qwen4}{QWEN4}{Qwen-3-4B}
\newacronym{gem12}{GEM12}{Gemma-3-12B}
\newacronym{phi04}{PHI04}{Phi-4}
\newacronym{gp020}{GP020}{GPT-OSS-20B}
\newacronym{grant}{GRANT}{Granite-4.0-H-Small}
\newacronym{devst}{DEVST}{Devstral Small}
\newacronym{gem27}{GEM27}{Gemma-3-27B}
\newacronym{qwenc}{QWENC}{Qwen-3 Coder}
\newacronym{llama}{LLAMA}{Llama-3.3}
\newacronym{gp120}{GP120}{GPT-OSS-120B}
\newacronym{gem31}{GEM31}{Gemma-4-31B}
\newacronym{gpt41}{GPT41}{GPT-4.1-2025-04-14}
\newacronym{loc}{LOC}{Lines of Code}
\newacronym{gt}{GT}{Ground Truth}
\newacronym{pypi}{PyPI}{Python Package Index}
\newacronym{npm}{NPM}{Node Package Manager}
\newacronym{api}{API}{Application Programming Interface}
\newacronym{tp}{TP}{True Positive}
\newacronym{fp}{FP}{False Positive}
\newacronym{tn}{TN}{True Negative}
\newacronym{fn}{FN}{False Negative}
\newacronym{oss}{OSS}{Open Source Software}

\def\BibTeX{{\rm B\kern-.05em{\sc i\kern-.025em b}\kern-.08em
T\kern-.1667em\lower.7ex\hbox{E}\kern-.125emX}}
\begin{document}

\title{An Evaluation of Large Language Models for Detection of Malicious Python Packages}

    \author{
        \IEEEauthorblockN{
        Ahmed Ryan\textsuperscript{1}\quad
        Ibrahim Khalil\textsuperscript{2}\quad
        Abdullah Al Jahid\textsuperscript{2}\quad
        Md Erfan\textsuperscript{1}\\
        Sungbin Park\textsuperscript{3}\quad
        Akond Ashfaque Ur Rahman\textsuperscript{4}\quad
        Md Rayhanur Rahman\textsuperscript{1}
        }
        \IEEEauthorblockA{
        \textsuperscript{1}University of Alabama, \{aryan9, merfan, mrahman87\}@crimson.ua.edu\\
        \textsuperscript{2}University of Dhaka, \{bsse1009, bsse1030\}@iit.du.ac.bd\\
        \textsuperscript{3}Auburn High School, tp081014@gmail.com\\
        \textsuperscript{4}Auburn University, azr0154@auburn.edu
        }
    }


    \maketitle

    \begin{abstract}
        Modern software development relies on open-source package repositories where developers share and reuse software components. 
        Attackers increasingly use these repositories to distribute malicious software packages, which embed code intended to harm the host systems that install them.
        \glspl{llm} are well-suited to automatic detection of such packages, given their ability to read and reason about code.
        It remains unclear whether LLMs can go beyond flagging a package as malicious and pinpoint the behaviors that make it so.
        To this end, we evaluate 13 LLMs on two tasks across a dataset of 4,070 PyPI packages, comprising 370 malicious and 3,700 benign samples.
        The first task is to detect whether a package is malicious or safe.
        The second task is to identify malicious indicators (i.e., the specific lines of code that make a package malicious), chosen from a set of 47 known indicator types grouped into 7 categories.
        We evaluate each \gls{llm} across five prompt strategies and three temperatures. 
        For the first task, \glspl{llm} show mean $F_1$ scores ranging from $0.40$ to $0.99$ across models: they correctly detect most malicious packages but frequently flag safe packages as malicious.
        For the second task, LLMs show a weighted $F_1$ score of $0.69$ for recognizing the type of malicious behavior but drop to $0.48$ when identifying the specific indicators responsible. 
        \glspl{llm} correctly identify indicators expressed as recognizable code patterns but miss indicators that require understanding the broader context or the author's intent. 
        Moreover, \glspl{llm} incorrectly report indicators that are not present in the package.
        We further evaluate the association between indicator identification performance and five factors: model size, context width, prompt strategy, temperature, and code complexity.
        Of these, only code complexity shows a meaningful association: longer packages are more difficult to analyze.
        Overall, we advocate that LLMs should be used for initial triage to flag suspicious packages for human review, not as tools that identify what a malicious package does.
        
    \end{abstract}


    \glsresetall

\section{Introduction}
\label{Introduction} 


Software packages play a pivotal role in modern software engineering, where developers use package repositories, such as \gls{pypi} for distributing the packages.
While these packages aid in software development, adversaries craft malicious packages, which embed code intended to harm the host systems that install them, and distribute those via the package repositories.
For example, on March 31, 2026 adversaries uploaded malicious code in a popular OSS package called `litellm'~\cite{attack:litellm} that harvested cloud credentials and API keys.
This Python package was downloaded 3 million times in one day.
The attack was estimated to impact 40,000 downloads~\cite{attack:litellm}.
As another example, in September 2025, adversaries injected malicious code to steal cryptocurrency into 18 widely used npm packages, which collectively account for over two billion weekly downloads~\cite{sygnia_npm_supply_chain_2025}.
According to one research study, the frequency of malicious packages increased by 156\% in one year~\cite{tosem25:laurie}.


The above discussion highlights the importance of detecting malicious packages, i.e., software packages in which adversaries intentionally embed harmful functionality to compromise the confidentiality, integrity, or availability of a system.  
The recurrence of security attacks via malicious software packages has garnered attention and initiatives in regards to securing \gls{oss} packages and ecosystems from industry, and academics~\cite{zahan2024s3c2summit202311industry, tosem25:laurie}.
Despite these efforts, malicious software packages are prevalent.  
For multiple factors, such as time pressure~\cite{linux:maintainers, perumachase2026:npm},  complexity in resolving third party dependencies~\cite{chainguard:package:pain}, and threat sophistication~\cite{perumachase2026:npm}, maintainers can miss harmful functionality in source code, which in turn leads to malicious packages being undetected.   


We use Figure~\ref{fig-intro} as an example in this regard. 
The figure presents a Python code snippet obtained from a package called \texttt{11Cent-999.0.4} which is indexed in the \texttt{pypi\_malregistry} dataset~\cite{guo2023empirical}. 
The code in \textit{line no 3} is an example of `Data Exfiltration'. 
This code snippet is an indicator, i.e., a code snippet that makes a package malicious. 
The malicious package has a total of 75 lines of code.  
If a user is not aware of the package being malicious, then the code snippet will reside in the user's source code and send sensitive directory data to the adversary. 


\vspace{-8pt}
\begin{listing}[h]
\centering
\begin{lstlisting}[
    language=Python,
    breaklines=true,
    basicstyle=\footnotesize\ttfamily,
    numbers=left,
    numberstyle=\tiny,
    xleftmargin=5pt,
]
ploads = {'/': ld0, '/root/': ld1, '/home/': ld2, '/root/.ssh/': ld3}
headers = {'content-type': 'application/x-www-form-urlencoded'}
requests.post('https://j0j0.xyz/lists/', data=ploads, verify=True, headers=headers)
\end{lstlisting}
\caption{An example of `Data Exfiltration', which is an indicator of a malicious Python package.}
\label{fig-intro}
\end{listing}
\vspace{-8pt}


One approach to aid maintainers can be usage of large language models (LLMs) because of their capabilities in secure software development tasks, such as penetration testing, fuzzing, vulnerability discovery, malware analysis, and code auditing~\cite{khare2025understanding, ullah2024llmsreliablyidentifyreason_p61, saha2025malaware_p19, zhang2025automatically_p20}. 
Recent works show the lack of automation as a challenge for practitioners when it comes to securing software packages, which makes the usage of LLMs a potential approach to automatically detect malicious packages.
A systematic empirical investigation can determine how accurately LLMs can detect malicious software packages in real-world ecosystems, how well they identify indicators, (i.e., coding patterns in Python source code that make a Python package malicious), what are the causes of detection performance, and which factors influence their detection performance.

\textbf{The goal of this paper is to help practitioners and researchers in understanding the capabilities of LLMs and how they perform in malicious package detection and indicator identification by conducting a systematic empirical evaluation of LLMs' performance. }
We investigate the following research questions:


\begin{itemize}

\item{\textbf{\textit{RQ1}}: \textit{How accurately can large language models (LLMs) detect malicious Python packages? How accurately can LLMs detect indicators of malicious Python packages?} }

\item{\textbf{\textit{RQ2}}: \textit{What are the root causes for LLMs' performance in detecting malicious Python packages and indicators of malicious Python packages?} }

\item{\textbf{\textit{RQ3}}: \textit{How are parameter size, context width, prompt strategy, temperature, and code complexity related with LLMs' performance in detecting indicators of malicious packages?} }

\end{itemize}

We conduct an empirical study on 13 \glspl{llm} based on two tasks i.e. (a) package detection and (b) category recognition and indicator identification to answer our research questions.
For $RQ_1$, we report precision, recall, and $F_1$ at three granularities namely, package,  category, and indicator. We also compare LLM performance across prompt strategies, temperatures, and \gls{llm} categories.  
For $RQ_2$, we partition every response of second task into eight mutually exclusive outcomes, rank the most-missed and most-hallucinated indicators, identify reasons behind outcomes, and group them by investigation effort.  
For $RQ_3$, we measure the association between $F_1$ and five factors (parameter size, context width, temperature, prompt strategy, and code complexity) using non-parametric statistics (Spearman's $\rho$ and the Kruskal-Wallis test) interpreted by effect size.


\textbf{\textit{Contributions}}: We list our contributions as follows:

\begin{itemize}[leftmargin=*]

\item{An evaluation of 13 \glspl{llm} across three granularities: package detection, category recognition, and indicator identification; }

\item{An eight-outcome taxonomy with investigation-effort grouping that exposes a systematic \glspl{llm} bias towards under-detecting context-dependent indicators and over-applying labels with recognizable syntax.} 

\item{An evaluation of how parameter size, context width, temperature, and prompt strategy are related with indicators of malicious Python packages.}

\end{itemize}    

    \section{Key Concepts}
    \label{sec:key-concepts}

    \noindent In this section, we present several key concepts.

    \textit{Malicious Indicator: }
    An observable signal in package metadata, source code, or runtime behavior that suggests malicious intent.
    Indicators can be coarse-grained (e.g., suspicious package names), fine-grained (e.g., \gls{api} call sequences, sensitive-source to untrusted-sink data flows).
    Indicators can even span statement and function levels, where individual indicators may be benign in isolation but malicious in combination.
    In Figure~\ref{fig-intro}, \textit{Data Exfiltration} indicator is present in line number three.

    \textit{Malicious Indicator Category: }
    A malicious indicator category is a high-level grouping of related malicious indicators based on common behavior.
    The taxonomy we adopt organizes the 47 indicators into 7 categories~\cite{ryan2025unveiling_p25}.
    For example, \textit{Typosquatting}, and \textit{Combosquatting} indicators belong to \textit{Metadata Manipulation} category.

    \textit{Detection, Recognition, and Identification: }
    \textit{Detection} is the verdict of whether a package is malicious or benign.
    \textit{Recognition} is the assignment of one or more of the seven high-level malicious-indicator \textit{categories} to a package (i.e., which kind of malicious behavior is present). 
    \textit{Identification} is the assignment of one or more of the 47 malicious indicators to a package (i.e., which malicious behavior is present).

    \textit{Weighted and Macro Average: }
    \label{paragraph:weighted-and-macro-average}
    A weighted average accounts for class frequency, reflecting overall \glspl{llm} performance across the dataset. 
    In contrast, a macro average treats all classes equally, highlighting performance on rare classes.

    \textit{Hallucination: }
    A phenomenon where \glspl{llm} generate false, fabricated, or logically inconsistent information while presenting it as factual truth~\cite{huang2025hallucination}.

    \section{Related Work}
    \label{sec:related}

    \noindent In this section, we present the related works.

    \textit{Malicious Package Detection: }
    Prior work has built datasets of real-world malicious packages across \gls{pypi}, \gls{npm}, and RubyGems~\cite{mehedi2025qut_p17, vu2022benchmark_p24, ohm2020backstabber_p29, zahan2024malwarebench_p33, ladisa2023hitchhiker_p34, vu2023bad_p36}. 
    Recent datasets have added statement-level indicator taxonomies~\cite{ryan2025unveiling_p25} and ATT\&CK-aligned labels~\cite{zhang2024tactics_p28}. 
    Empirical studies characterize malicious-package behavior and the high false-positive rates of existing detectors~\cite{ladisa2023hitchhiker_p34, zhou2024large_p37, guo2023empirical, vu2022benchmark_p24, vu2023bad_p36}.
    Detection techniques comprises metadata and code-metric features~\cite{Halder2024MaliciousPD_p47, yan2025pypimaldet_p07, samaana2025machine_p22, Ladisa_2023_p40}, installation-script anomaly detection~\cite{liang2025detecting_p10, liang2023needle_p39, Liang2021MaliciousPL_p44}, source-versus-artifact differential analysis~\cite{shariffdeen2025detecting_p14, vu2021lastpymile_p31, froh2023differential_p32, Scalco2022OnTF_p45}, dataflow, call-graph, slicing, and bytecode analysis~\cite{yu2024maltracker_p35, wang2025malpacdetector_p12, 10298343_p41, Ladisa2022TowardsTD_p46}, dynamic sandboxing and kernel-level tracing~\cite{cohen2025javasith_p18, zhang2025killing_p23, nguyen2024analysis_p30, Huang2024DONAPIMN_p43, mehedi2026dysec_p13, tan2026operational_p08}, and traditional, graph, and ensemble learning over these features~\cite{singh2020detection_p21, Sejfia2022PracticalAD_p48, huangprofmal_p03, iqbal2025pypiguard_p15}.

    \textit{\gls{llm}-based Detection: }
    \glspl{llm} are increasingly used as classifiers for malicious code, often with few-shot or retrieval-augmented prompting~\cite{ibiyo2025detecting_p05, sun20241+_p27}. 
    \glspl{llm} are also integrated into static pipelines to flag sensitive \glspl{api}, refine features, or guide slicing~\cite{gao2025malguard_p06, huang2024spiderscan_p26, nguyen2025taint_p02}.
    Recent agentic works include multi-agent dissection~\cite{toda2025chase_p09}, agent workflows with suspicious-\gls{api} knowledge and external tools~\cite{pypiline2026}, knowledge mining~\cite{gordian2026}, and adaptive registry-to-enterprise detectors~\cite{montaruli2025one_p01}.
    \glspl{llm} also summarize malicious behavior~\cite{saha2025malaware_p19} and generate YARA/Semgrep rules~\cite{zhang2025automatically_p20}.
    MalLoc~\cite{malloc2025} is the closest to our work. 
    This work localizes malicious payloads with \glspl{llm}, but targets Android and does not compare fine-grained localization against coarse-grained package detection.

    \textit{Evaluation of \glspl{llm} on Security: } 
    Comparative studies report variation in \gls{llm} performance with frequent false positives and non-determinism~\cite{wyssevaluating_p11, dai2025rethinkingevaluationsecurecode_p53, han2025llmsagentscomparativeevaluation_p54, ullah2024llmsreliablyidentifyreason_p61}.
    The studies also find that \glspl{llm} analyze isolated logic but fail when required to track data, and logical dependencies across multiple functions, files, or systems~\cite{Lin2025FromLT_p58, khare2025understanding, Huynh2025DetectingCV_p64}.
    Agent and benchmark studies report gaps in workflow evaluation in real-world environment~\cite{han2025llmsagentscomparativeevaluation_p54, lee2025secbenchautomatedbenchmarkingllm_p55, yildiz2025benchmarkingllmsllmbasedagents_p59, Wang2025CanLR_p65, sastbench2026, tosss2026}. 
    The studies further report that using aggregate scores to evaluate \glspl{llm} hide performance flaws~\cite{seclens2026}.
    \glspl{llm} identify threats but generate high volumes of false positives and require calibration of sensitivity rather than better core detection capabilities~\cite{opensec2026}.
    Other studies report that \gls{llm} effectiveness is task-dependent and often below expert-designed or hybrid methods~\cite{Zhang2023HowWD_p50, Licorish2025ComparingHA_p51, rondanini2025malwaredetectionedgelightweight_p57}.
    A study related to our work, evaluates whether \glspl{llm} can detect malicious updates to \gls{npm} packages~\cite{wyssevaluating_p11}, but that study does not separate package-level detection from indicator-level identification.

    Existing evaluations measure a single granularity (usually a package-level verdict) and rarely test whether \glspl{llm} can identify the specific behaviors responsible for that verdict.
    We evaluate 13 \glspl{llm} at both the package and indicator levels over a 47-indicator statement-level taxonomy~\cite{ryan2025unveiling_p25} and analyze the failure patterns behind the gap.

    \section{Methodology}

    \noindent In this section, we present the methodology.
    
    \subsection{Task Formulation}
    \label{sec:task_formulation}

    We formulate two classification tasks to evaluate \glspl{llm} capabilities for malicious package analysis.
    \textbf{Task 1 (Malicious Package Detection): }
    A binary classification task designed to determine whether an \gls{llm} can distinguish between benign and malicious packages based on source code.
    \textbf{Task 2 (Category Recognition and Indicator Identification): }
    A multi-label classification task designed to assess whether an \gls{llm} can recognize the coarse-grained categories in the code and identify fine-grained indicators.

    \subsection{Package Collection}

    We collect malicious and benign packages to carry out two tasks.
    Task-1 requires both malicious and benign packages whereas Task-2 requires only malicious packages.
    With 717,280 total packages in PyPI \cite{pypistats} and over 10,000 recorded malicious packages in \textit{pypi\_malregistry} (last accessed June 24, 2026), we estimate the ecosystem's benign-to-malicious ratio to be $\approx 73:1$
    Since experimenting with such a ratio is resource-intensive, we maintain a 10:1 benign-to-malicious package ratio for Task-1 following the methodology established in~\cite{qing2024rapier}.
    The resulting ratio remains computationally feasible while ensuring that benign packages outnumber malicious packages to reflect a skewed ecosystem.

    \textit{Malicious Package Collection: }
    For Task 2, we require malicious Python packages with statement-level labels instead of a single label for the entire package.
    Statement-level labels pinpoint the exact \gls{loc} where malicious behavior is present.
    To the best of our knowledge, only one study has constructed such a dataset with statement-level labels~\cite{ryan2025unveiling_p25}.
    The dataset contains \textbf{370 malicious packages} (a package comprises one or more files). 
    The packages have a total of 385 files and 2,962 statement-level labels annotated with a taxonomy of 47 indicators.
    For Task 1, we do not require the statement-level labels, single label for the entire package is sufficient.
    We collect all the malicious packages of this dataset.

    \textit{Benign Package Collection: }
    For Task 1, we require benign packages.
    We initially collect 9,127 benign packages from the dataset provided by the \textit{MalGuard} study \cite{gao2025malguard_p06}.
    We analyze the dataset and find a heavy-tailed distribution, with a mean of 134,148 \gls{loc} but a median of only 288.
    Malicious packages are usually small, as in our malicious packages, the median \gls{loc} is 36.
    To align with the size of the malicious packages and maintain compatibility with \glspl{llm}' context windows, we filter the benign packages with a minimum of 20 \gls{loc}. 
    Then we rank the resulting packages by \gls{loc} in ascending order, and choose the top \textbf{3,700 benign packages}.
    In the resulting packages, the highest \gls{loc} we observe is 260.

    \subsection{Model Selection}

    We categorize \glspl{llm} by availability (open-source vs. proprietary) and specialization (general vs. coder).
    This setup helps to understand the performance of open-source \glspl{llm} against commercial \glspl{llm} and whether specialized code training outperforms general reasoning.
    We select 13 \glspl{llm} from 7 different providers to cover variety, as shown in Table \ref{tab:model_selection}.

\begin{table}[htbp]
    \vspace{-5pt}
    \scriptsize
    \centering
    \caption{Overview of Selected Large Language Models}
    \vspace{-5pt}
    \label{tab:model_selection}
    \setlength{\tabcolsep}{5pt}
    \begin{tabular}{lrrcc}
        \toprule
        \textbf{Model (Id)} & \textbf{Params (B)} & \textbf{Context (T)} & \textbf{Access} & \textbf{Type} \\
        \midrule
        \gls{gem04} & 4 & 128k & OS & G \\
        \gls{qwen4} & 4 & 256k & OS & G \\
        \gls{gem12} & 12 & 128k & OS & G \\
        \gls{phi04} & 15 & 16k & OS & G \\
        \gls{gp020} & 20 & 128k & OS & G \\
        \gls{grant} & 32 & 131k & OS & G \\
        \gls{devst} & 24 & 128k & OS & X \\
        \gls{gem27} & 27 & 128k & OS & G \\
        \gls{qwenc} & 30 & 256k & OS & X \\
        \gls{llama} & 70 & 128k & OS & G \\
        \gls{gp120} & 120 & 128k & OS & G \\
        \gls{gem31} & 33 & 262k & OS & G \\
        \gls{gpt41} & Unknown & 1000k & P & G \\
        \bottomrule
    \end{tabular}
    \\[3pt]
    \raggedright \centering \scriptsize \textit{Note: B=Billions, T=Tokens, OS=Open-Source, P=Proprietary, G=General, X=Coder. Model Ids denote these \glspl{llm} throughout the paper.}
    \vspace{-10pt}
\end{table}

    \subsection{Configuration}

    \textit{Prompt Strategy: } 
    Prompt strategy directs the reasoning process.
    We employ five strategies by increasing complexity and computational cost, to measure how the prompts affect performance.
    Each successive strategy demands more resources due to larger input contexts, increased token generation, or multiple model iterations. 
    \textit{\gls{zs}} prompting requires the model to generate a response based solely on pre-trained knowledge without provided examples~\cite{brown2020language}.
    \textit{Few-Shot} prompting advances this by including a small number of input-output examples in the context window to demonstrate the desired format or logic~\cite{brown2020language}. 
    \textit{Chain-of-Thought (CoT)} introduces linear reasoning, prompting the model to output intermediate reasoning steps before arriving at a final answer~\cite{wei2022chain}. 
    \textit{Self-Consistency} builds upon CoT through parallel consensus, generating multiple independent reasoning paths and selecting the most frequent final answer~\cite{wang2022self}. 
    \textit{Tree-of-Thoughts (ToT)} utilizes branching logic, enabling the model to explore, evaluate, and backtrack across multiple reasoning steps to solve complex problems~\cite{yao2024tree}.

    \textit{Temperature: }
    We employ three temperature ($\tau$) to measure how determinism and diversity affects performance. At $\tau = 0$, the model deterministically selects the token with highest probability, which can increase repetition~\cite{holtzman2020curious}. 
    At $\tau = 0.5$ and $\tau = 1$, temperature scaling adjusts output diversity: lower temperature provides coherence, higher temperature provides lexical diversity~\cite{ackley1985learning,holtzman2020curious}.

    \subsection{Sampling Strategy}

    Our experiment involves 13 \glspl{llm}, 5 prompt strategies, and 3 temperatures.
    We use different sampling strategies for the two tasks, where Task-1 includes both malicious and benign packages, and Task-2 includes only malicious packages.

    \textit{Task-1 (Binary Classification): }
    We create a 4,070-package corpus (370 malicious + 3,700 benign).
    A run over such a large corpus would require 793,650 responses (4,070 packages $\times$ 13 models $\times$ 5 prompts $\times$ 3 temperatures), which is computationally expensive.
    Hence, we sample $2.5\%$ from both 370 malicious and 3,700 benign packages, resulting 10 malicious packages and 93 benign packages.
    We take one file from each package, and use a fixed seed so that every configuration (prompt, temperature) receives identical packages.
    We evaluate the open-source model at three temperatures, but we evaluate the proprietary \gls{llm} (GPT-4.1) at two temperatures instead of three to reduce expense.
    In total, we generate 19,570 responses: $103$ files $\times 12$ \glspl{llm} $\times 5$ prompts $\times 3$ temperatures for open-source and $103$ files $\times 1$ \gls{llm} $\times 5$ prompts $\times 2$ temperatures for proprietary.

    \textit{Task-2 (Multi-Label Classification): }
    We use 370 malicious packages (385 files) for every configuration (prompt, temperature).
    We generate 73,150 responses: $385$ files $\times 12$ \glspl{llm} $\times 5$ prompts $\times 3$ temperatures for open-source and $385$ files $\times 1$ \gls{llm} $\times 5$ prompts $\times 2$ temperatures for proprietary.

    \subsection{Task Evaluation}
    \label{sec:method-eval}

    We evaluate both tasks using precision, recall, and $F_1$ score, computed from \gls{llm} responses against ground-truth labels.

    \textit{Task-1 (Binary Classification): }
    We compute precision, recall, and $F_1$ score across all sampled packages. 
    We aggregate these results to report performance grouped by \gls{llm}, prompt strategy, and temperature.

    \textit{Task-2 (Multi-Label Classification): }
    We evaluate this task at two granularities: the 47 fine-grained indicators and their 7 coarse-grained categories. 
    For category-level evaluation, we map the indicators to their corresponding categories. 
    At both granularities, if a label is present in both the ground truth and the prediction, we count it as a success, regardless of the frequency of the labels. 
    We summarize these per-label scores using macro and weighted averages (defined in Section~\ref{paragraph:weighted-and-macro-average}).

    \subsection{Analysis on Incorrect Indicator Identification}
    \label{subsection:methodology-error-analysis}

    To characterize how the \glspl{llm} incorrectly identified indicators ($RQ_2$), we compare the predicted label set of all \gls{llm} responses against the ground-truth label set.

    \textit{Outcome Taxonomy: }
    We create an outcome taxonomy from the eight possible combinations of the presence ($>0$) or absence ($0$) of \gls{tp}, \gls{fp}, and \gls{fn} (see Table~]\ref{tab:response_outcomes}).
    We classify each \gls{llm} response into one of eight mutually exclusive outcomes.

    \begin{table}[htbp]
        \vspace{-5pt}
        \centering
        \caption{Outcome Taxonomy}
        \vspace{-5pt}
        \label{tab:response_outcomes}
        \setlength{\tabcolsep}{10pt}
        \begin{tabular}{lcccl}
            \toprule
            \textbf{Outcome} & \textbf{TP} & \textbf{FN} & \textbf{FP} & \textbf{Status} \\
            \midrule
            correct\_clean\textsuperscript{*} & $0$ & $0$ & $0$ & Success \\
            correct\_hit & $>0$ & $0$ & $0$ & Success \\
            under\_detection & $>0$ & $>0$ & $0$ & Partial Success \\
            over\_detection & $>0$ & $0$ & $>0$ & Partial Success \\
            mixed & $>0$ & $>0$ & $>0$ & Partial Success \\
            false\_alarm\textsuperscript{*} & $0$ & $0$ & $>0$ & Failure \\
            silent\_miss & $0$ & $>0$ & $0$ & Failure \\
            mismatch & $0$ & $>0$ & $>0$ & Failure \\
            \bottomrule
        \end{tabular}
        \vspace{-5pt}
    \end{table}

    We interpret Table~\ref{tab:response_outcomes}.
    \textit{correct\_clean:} \gls{llm} correctly identifies a benign package.
    \textit{correct\_hit:} \gls{llm} correctly identifies all actual indicators present in the package and does not incorrectly identify any indicator that is not present in the package.
    \textit{under\_detection:} \gls{llm} correctly identifies some actual indicators present in the package but does not identify some actual indicators that are present in the package.
    \textit{over\_detection:} \gls{llm} correctly identifies all actual indicators present in the package and incorrectly identifies some indicators that are not present in the package.
    \textit{mixed:} \gls{llm} correctly identifies some actual indicators present in the package, does not identify some actual indicators present in the package, and incorrectly identifies some indicators that are not present in the package.
    \textit{false\_alarm:} In a benign package, \gls{llm} incorrectly identifies indicator(s) that are not present.
    \textit{silent\_miss:} \gls{llm} does not identify any actual indicators present in the package.
    \gls{llm} does not incorrectly identify any indicator that is not present in the package. 
    \textit{mismatch:} \gls{llm} does not identify any actual indicator present in the package and incorrectly identifies some indicators that are not present in the package.
    As every response in Task-2 comes from malicious packages, \textit{correct\_clean} and \textit{false\_alarm} are zero by construction.

    \textit{Miss, Hallucination, and Mismatch Analysis: }
    For each indicator, we analyze actual indicators in the package that \glspl{llm} do not identify (referred to as miss). 
    We also analyze indicators identified by \glspl{llm} that are not present in the package (referred to as hallucination).
    We rank indicators by these measures to identify the most-missed and most-hallucinated techniques and map each to its corresponding category.

    \textit{Qualitative and Configuration Analysis: }
    We inspect \textit{mismatch} outcomes of all the 13 \glspl{llm} to understand their causes, and we show representative examples for each cause.
    Finally, we compare the outcome distribution across configurations (prompt strategy, temperature) and report the findings.

    \textit{Classification based on Investigation Effort: }
    We categorize outcomes by subsequent investigation effort required by human analysts after \glspl{llm} triage, ranging from \textit{low} (\textit{over\_detection}) through \textit{medium} (\textit{under\_detection}) and \textit{high} (\textit{mixed}) to \textit{critical} (\textit{mismatch}, \textit{silent\_miss}), where effort increases with the combination of false positives and missed indicators.

    \subsection{Correlation Analysis}
    \label{sec:method-correlation}

    To evaluate how different factors influence indicator identification by \glspl{llm}, we measure the correlation between the $F_1$ score and five factors: parameter size, context width, temperature, prompt strategy, and code complexity.
    For context width, the sample size is the count of \glspl{llm} $(N=13)$.
    For parameter size, we exclude the proprietary \gls{llm} \gls{gpt41} as we do not know its exact parameter size $(N=12)$.
    For code complexity, the sample size is the count of malicious packages $(N=370)$. 
    For temperature and prompt strategy, we evaluate all responses ($N = 50{,}892$) that was successfully parsed. 
    The remaining responses out of the initial $73{,}150$ were excluded due to \gls{api} timeouts, formatting issues, or execution errors.

    As the $F_1$ score distribution is bounded and non-normal, the analysis relies on non-parametric statistical testing.
    Spearman's rank correlation ($\rho$) evaluates continuous and ordinal factors (parameter size, context width, temperature, and code complexity).
    The Kruskal-Wallis $H$ test evaluates the nominal prompt strategy variable, deriving the $\eta^2 = H / (N - 1)$ effect size.
    When the sample size is large ($N \approx 50{,}000$), a small difference results in a statistically significant p-value, even if the correlation is tiny.
    To determine if the results are practically significant, we rely on effect sizes, and treat p-value as secondary.
    We assess correlational strength using absolute rank correlations ($|\rho|$), classifying coefficients $<0.10$ as negligible and $>0.50$ as large.
    Concurrently, we quantify the proportion of explained variance using eta-squared ($\eta^2$), defining $<0.01$ as negligible and $>0.14$ as large~\cite{cohen1988statistical}.

    \section{Findings}

    \noindent In this section, we present the findings.

    \subsection{Findings of $RQ_1$}
    \label{section:findings-rq1}

    In this section, we present the findings of $RQ_1$.

    \subsubsection{Capability of LLMs in Detection}

    We present the $F_1$ scores of \glspl{llm} across all configurations for malicious package detection in  Table~\ref{tab:package_detection}.

    \textit{Overall Detection Performance: }
    \label{overall-detection-performance}
    Across all configurations, mean $F_1$ scores for the \glspl{llm} range from $0.40$ to $0.99$.
    We calculate the aggregated mean for each prompt strategy by averaging \glspl{llm}' performance with that prompt across all three temperatures.
    \gls{zs} shows the highest aggregated mean ($F_1=0.74$), derived from its mean scores at temperatures 0 ($F_1=0.75$), 0.5 ($F_1=0.72$), and 1 ($F_1=0.76$).
    \gls{zs} is followed by \gls{tot} at $0.67$, \gls{cot} at $0.58$, \gls{sc} at $0.58$, and \gls{fs} at $0.23$.
    Similarly, we calculate aggregated mean for each temperature by averaging \glspl{llm}' performance with that temperature across all five prompts.
    Temperature 0 produces an aggregated mean of $0.56$ (derived from $0.75$, $0.25$, $0.62$, $0.68$, and $0.52$), while temperature 1 and 0.5 shows $0.58$ and $0.53$, respectively.

\begin{table}[htbp]
    \centering
    \scriptsize
    \caption{Package Detection Performance}
    \vspace{-5pt}
    \label{tab:package_detection}
    \setlength{\tabcolsep}{1.75pt}
    \begin{tabular}{lccccccccccccccccc}
        \toprule
        \textbf{Prompt} & \multicolumn{3}{c}{\textbf{ZS}} & \multicolumn{3}{c}{\textbf{FS}} & \multicolumn{3}{c}{\textbf{CoT}} & \multicolumn{3}{c}{\textbf{ToT}} & \multicolumn{3}{c}{\textbf{SC}} & Mean & Max \\
        \cmidrule(lr){2-4} \cmidrule(lr){5-7} \cmidrule(lr){8-10} \cmidrule(lr){11-13} \cmidrule(lr){14-16}
        \textbf{T} & 0 & 0.5 & 1 & 0 & 0.5 & 1 & 0 & 0.5 & 1 & 0 & 0.5 & 1 & 0 & 0.5 & 1 & & \\
        \midrule
        GPT41 & \textbf{1} & - & \textbf{1} & \textbf{1} & - & \textbf{1} & \textbf{1} & - & \textbf{1} & \textbf{1} & - & \textbf{1} & \textbf{.95} & - & .95 & \textbf{.99} & \textbf{1} \\
        LLAMA & .92 & .92 & .92 & .15 & .15 & .15 & .75 & .75 & .54 & .89 & .92 & .92 & .80 & .80 & .91 & .70 & .92 \\
        QWEN4 & .82 & .82 & .82 & \underline{.14} & \underline{.14} & \underline{.14} & .70 & .56 & .55 & .86 & \textbf{1} & \textbf{1} & .92 & \textbf{.92} & .92 & .69 & \textbf{1} \\
        GP120 & .86 & .80 & .86 & .17 & .17 & .18 & .80 & .67 & .80 & .86 & .86 & .89 & .71 & .71 & .67 & .67 & .89 \\
        GP020 & .83 & .83 & .77 & .18 & .21 & .17 & .29 & .29 & .36 & .92 & .92 & .91 & .80 & .83 & .83 & .61 & .92 \\
        QWENC & .92 & .92 & .92 & .15 & .15 & .15 & .67 & .58 & .44 & .83 & .86 & .75 & \underline{.00} & .67 & \textbf{1} & .60 & \textbf{1} \\
        GEM31 & \textbf{1} & \textbf{1} & \textbf{1} & .50 & \textbf{.50} & .52 & \textbf{1} & \textbf{1} & \textbf{1} & \underline{.15} & \underline{.15} & \underline{.15} & .30 & .30 & .30 & .59 & \textbf{1} \\
        PHI04 & \underline{.15} & \underline{.14} & \underline{.15} & .15 & \underline{.14} & .15 & .55 & .52 & .39 & .82 & .75 & .75 & .86 & .86 & .92 & .49 & .92 \\
        GEM27 & .80 & .80 & .92 & .18 & .19 & .18 & .54 & .52 & .74 & .26 & .27 & .41 & \underline{.00} & \underline{.00} & .67 & .43 & .92 \\
        GRANT & .92 & .92 & .92 & \underline{.14} & .15 & \underline{.14} & .77 & .77 & .43 & .25 & .21 & .20 & .15 & .20 & \underline{.19} & .43 & .92 \\
        DEVST & .29 & .27 & .32 & .16 & .15 & \underline{.14} & \underline{.14} & \underline{.21} & \underline{.19} & .86 & .75 & .80 & .67 & .63 & .80 & .43 & .86 \\
        GEM12 & .71 & .67 & .67 & .22 & .17 & .17 & .35 & .36 & .37 & .62 & .50 & .50 & .24 & .29 & .25 & .41 & \underline{.71} \\
        GEM04 & .55 & .57 & .60 & .17 & .18 & .16 & .47 & .33 & .48 & .53 & .75 & .25 & .39 & .27 & .33 & \underline{.40} & .75 \\
        \midrule
        Mean & .75 & .72 & .76 & .25 & .19 & .25 & .62 & .55 & .56 & .68 & .66 & .66 & .52 & .54 & .67 & & \\
        Best & 1 & 1 & 1 & 1 & .50 & 1 & 1 & 1 & 1 & 1 & 1 & 1 & .95 & .92 & 1 & & \\
        \bottomrule
    \end{tabular}
    \\[3pt]
    \raggedright \centering \scriptsize Note: Precision (P), Recall (R), $F_1$ Score ($F_1$). Combination of Precision and Recall that generated highest $F_1$ (Best). Zero Shot (ZS), Few Shot (FS), CoT (Chain of Thought), ToT (Tree of Thought), SC (Self Consistency).
    \vspace{-20pt}
\end{table}

    The maximum $F_1$ scores range from $0.71$ to $1.00$.
    Four \glspl{llm} show maximum score ($F_1 = 1.00$) in at least one configuration.
    Furthermore, every prompt strategy and temperature setting reaches the maximum score at least once.
    
    \textit{Impact of Configuration Selection: }
    An optimal configuration as a combination of prompt strategy and temperature that produces the maximum $F_1$ score for a given \gls{llm}.
    Evaluated under the optimal configuration, 9 of the 13 \glspl{llm} show $F1>0.90$.
    Four of these \glspl{llm} (e.g., \gls{gpt41}, \gls{gem31}, \gls{qwen4}, and \gls{qwenc}) show $F1=1.0$, detecting all malicious packages with zero incorrect predictions.
    However, the $F_1$ scores depend on configuration.
    As mentioned earlier, $F_1$ scores fluctuated across the 15 configurations, with scores ranging from $0.00$ (\gls{qwenc} at \gls{sc}, T=0) to $1.0$ (\gls{qwenc} at \gls{sc}, T=1).
    As the same \gls{llm} can show such difference in $F_1$ score ($\Delta=1.0$) based on configuration, reporting an \gls{llm}'s $F_1$ score is meaningless unless the exact configuration is also defined.

    \textit{Precision and Recall Analysis: }
    In the context of malicious package detection, recall is the proportion of actual malicious packages the \gls{llm} detects and precision is the proportion of detected packages that are actual malicious packages.
    Across all configurations, \glspl{llm} maximize recall at the expense of precision.
    Few actual malicious packages remain undetected by \glspl{llm}, 1.6\% of configurations result in a recall below 0.5.
    On the contrary, \glspl{llm} produce numerous false positives, 51\% of configurations show precision below 0.5.
    The results demonstrate that most of \glspl{llm} incorrect predictions involve detecting benign packages as malicious.

    \textit{Impact of Prompt Strategy and Temperature: }
    Aggregated mean $F_1$ scores (defined and calculated in Section~\ref{overall-detection-performance}) show that prompt strategy impacts $F_1$ score more than temperature.
    Temperatures show a minor variance $\approx 0.05$ (ranging from $0.53$ to $0.58$).
    Prompt strategy show more variance $\approx 0.51$ (ranging from $0.23$ to $0.74$).
    The variances indicate that prompt strategy impacts the $F_1$ score more than temperature by $\approx 10$ times.

    \textit{Limitation of Few-Shot Prompt Strategy: }
    While \gls{zs} shows the highest aggregated mean $F_1=0.74$, \gls{fs} shows the lowest aggregated mean $F_1=0.23$.
    For context, if a \gls{llm} classifies every package as malicious, it will show $F_1$ score of $0.14$.
    Under \gls{fs} strategy, a single outlier \gls{gpt41} shows $F_1=1$, but all the other \glspl{llm} show $F_1$ scores ranging from $0.14$ to $0.52$.

    \textit{Coder vs General Models: }
    The 2 code-specialized \glspl{llm} (\gls{qwenc} and \gls{devst}) show aggregated mean $F_1=0.51$, while the 11 general-purpose \glspl{llm} perform slightly better with aggregated mean $F_1=0.59$.
    Both groups reach a maximum $F_1=1.0$.
    For our dataset, specialized code training offers no performance gain in malicious package detection.

    \subsubsection{Capability of LLMs in Recognition and Identification}

    In this section, we evaluate the performance of \glspl{llm} in category recognition and indicator identification.

    \textit{Overall Recognition and Identification Performance: }
    Table~\ref{tab:overall_performance_malicious_indicator_and_categories} shows macro and weighted precision, recall, and $F_1$.
    In category recognition, \glspl{llm} show a weighted $F_1=0.69$ (Recall $0.63$), leaving $37\%$ of categories unrecognized.
    The macro $F_1=0.67$ (Precision 0.79) indicates more false positives in categories with lower support compared to other categories.
    In indicator identification, \glspl{llm} show a weighted $F_1=0.48$ (Precision $0.70$, Recall $0.41$), leaving $59\%$ of indicators unidentified.
    The macro $F_1=0.35$ (Precision $0.43$, Recall $0.39$) indicates $61\%$ of indicators with lower support remain unidentified. 

    \begin{table}[htbp]
        \centering
        \caption{Recognition and Identification Performance}
        \vspace{-5pt}
        \label{tab:overall_performance_malicious_indicator_and_categories}
        \setlength{\tabcolsep}{5pt}
        \begin{tabular}{llcccccc}
            \toprule
            & & \multicolumn{3}{c}{\textbf{Mean}} & \multicolumn{3}{c}{\textbf{Best}} \\
            \cmidrule(lr){3-5} \cmidrule(lr){6-8}
            \textbf{} & \textbf{Average} & \textbf{P} & \textbf{R} & \textbf{$F_1$} & \textbf{P} & \textbf{R} & \textbf{$F_1$} \\
            \midrule
            Category Recognition & Macro & .79 & .65 & .67 & .79 & .71 & .71 \\
            & Weighted & .85 & .62 & .69 & .85 & .70 & .75 \\
            \midrule
            Indicator Identification & Macro & .43 & .39 & .35 & .45 & .46 & .40 \\
            & Weighted & .70 & .41 & .48 & .69 & .47 & .52 \\
            \bottomrule
        \end{tabular}
        \\[3pt]
        \raggedright \centering \scriptsize Note: Precision (P), Recall (R), $F_1$ Score ($F_1$), Combination of Precision and Recall that generated highest $F_1$ (Best).
        \vspace{-5pt}
    \end{table}

    \textit{Category Recognition Performance: }
    Across six of the seven categories, \glspl{llm} show precision scores ($0.71$ to $0.94$) that exceed recall scores ($0.38$ to $0.85$).
    This result demonstrates that \glspl{llm} leave actual categories unrecognized but are correct when they make a prediction.
    \textit{Network Operations} is the only exception.
    For this category, recall ($0.88$) exceeds precision ($0.34$), indicating that $66\%$ of \glspl{llm} predictions in this category are false positives.

    \begin{table}[htbp]
        \vspace{-5pt}
        \centering
        \caption{Category Recognition Performance}
        \vspace{-5pt}
        \label{tab:category-analysis-formatted}
        \setlength{\tabcolsep}{3.75pt}
        \begin{tabular}{lccccccc}
            \toprule
            & & \multicolumn{3}{c}{\textbf{Mean}} & \multicolumn{3}{c}{\textbf{Best}} \\
            \cmidrule(lr){3-5} \cmidrule(lr){6-8}
            \textbf{Category} & \textbf{Support} & \textbf{P} & \textbf{R} & \textbf{$F_1$} & \textbf{P} & \textbf{R} & \textbf{$F_1$} \\
            \midrule
            Execution Stage (ES) & 41,816 & .87 & .69 & .77 & .86 & .73 & .79 \\
            Execution Mechanism (EM) & 41,431 & \textbf{.94} & \textbf{.85} & \textbf{.89} & .93 & \textbf{.92} & \textbf{.92} \\
            Defense Evasion (DE) & 40,721 & \textbf{.94} & \underline{.38} & .55 & \textbf{.94} & .54 & .69 \\
            Metadata Manipulation (MM) & 35,318 & .87 & .46 & .60 & .86 & \underline{.52} & .65 \\
            System Impact (SI) & 15,210 & .71 & .69 & .70 & .74 & .76 & .75 \\
            Network Operations (NO) & 11,312 & \underline{.34} & \textbf{.88} & \underline{.49} & \underline{.33} & .90 & \underline{.48} \\
            Exfiltration (EX) & 9,038 & .85 & .56 & .67 & .84 & .61 & .71 \\
            \bottomrule
        \end{tabular}
        \\[3pt]
        \raggedright \centering \scriptsize Note: Precision (P), Recall (R), $F_1$ Score ($F_1$), Combination of Precision and Recall that generated highest $F_1$ (Best). \textbf{Bold} = highest value in column, \underline{underline} = lowest value in column. Table sorted in descending order by Support.
        \vspace{-5pt}
    \end{table}

    Mean $F_1$ score does not depend on sample size.
    \textit{Execution Mechanism} (Support $41,431$) show a mean $F_1$ of $0.89$. \textit{Defense Evasion} has similar support ($40,721$) but shows a mean $F_1$ of $0.55$.
    \textit{Defense Evasion} shows the highest variance between mean and optimal configurations.
    Applying optimal configuration increases recall by $0.16$ (from $0.38$ to $0.54$) and the $F_1$ score by $0.14$ (from 0.55 to 0.69) while maintaining a precision of $0.94$. 
    \textit{System Impact} has the most balanced precision-recall mean score. 
    The mean precision ($0.71$) and mean recall ($0.69$) differ by $0.02$.
    On the contrary, \textit{Defense Evasion} and \textit{Metadata Manipulation} show mean precision-recall differences of $0.56$ and $0.41$, respectively. 
    When the \glspl{llm} predict these categories, the predictions are almost always correct.
    However, \glspl{llm} leave the majority of actual categories unrecognized.

    \textit{Indicator Identification Performance: }
    We present the identification performance of each indicator in Table~\ref{tab:indicator_identification_performance}. 
    For the indicators with highest mean $F_1$ scores (e.g. \textit{Shell Command Execution} and \textit{Hidden Code Execution}), \glspl{llm} are correct in 86\% and 92\% of their predictions, respectively.
    Even in these cases, \glspl{llm} leave approximately one-third of the actual indicators unidentified, indicating low recall.

    \begin{table}[htbp]
        \vspace{-5pt}
        \centering
        \caption{Indicator Identification Performance}
        \vspace{-5pt}
        \label{tab:indicator_identification_performance}
        \setlength{\tabcolsep}{4pt}
        \scriptsize
        \begin{tabular}{l r ccc ccc}
            \toprule
            & & \multicolumn{3}{c}{\textbf{Mean}} & \multicolumn{3}{c}{\textbf{Best}} \\
            \cmidrule(lr){3-5} \cmidrule(lr){6-8}
            \textbf{Indicator (Category Codes)} & \textbf{Support} & \textbf{P} & \textbf{R} & \textbf{$F_1$} & \textbf{P} & \textbf{R} & \textbf{$F_1$} \\
            \midrule
            Suspicious Author Identity (MM) & 33223 & \textbf{.90} & .40 & .55 & \textbf{.88} & .43 & .58 \\
            Shell Command Execution (EM) & 30340 & .86 & \textbf{.71} & \textbf{.78} & .84 & \textbf{.79} & \textbf{.81} \\
            Error Suppression (DE) & 28908 & .85 & .19 & .31 & .85 & .28 & .42 \\
            Encoding-Based Obfuscation (DE) & 27906 & .89 & .30 & .45 & .87 & .39 & .53 \\
            Hidden Code Execution (EM) & 24160 & \textbf{.92} & .66 & \textbf{.77} & \textbf{.88} & .68 & \textbf{.77} \\
            Install-Time Execution (ES) & 23668 & .51 & .52 & .51 & .51 & .61 & .56 \\
            Decoy Functionality (MM) & 23340 & .66 & .12 & .20 & .67 & .18 & .28 \\
            Lifecycle Hook Hijack (ES) & 15540 & .84 & .45 & .59 & .82 & .48 & .61 \\
            System Info Reconnaissance (SI) & 12799 & .87 & .49 & .63 & .85 & .56 & .67 \\
            Dynamic Evaluation (EM) & 8539 & .63 & \textbf{.78} & \textbf{.70} & .60 & \textbf{.82} & .69 \\
            Data Exfiltration (EX) & 8530 & .82 & .35 & .49 & .85 & .42 & .56 \\
            Conditional Payload Trigger (EM) & 7742 & .70 & .23 & .34 & .66 & .29 & .40 \\
            Typosquatting (MM) & 5973 & .34 & .25 & .29 & .34 & .32 & .33 \\
            Dynamic Module Import (EM) & 5900 & .21 & \underline{.03} & \underline{.05} & .23 & \underline{.04} & .07 \\
            Suspicious Connection (NO) & 5686 & .25 & .34 & .29 & .23 & .38 & .29 \\
            Arbitrary File Write (SI) & 5289 & .36 & .56 & .44 & .41 & .65 & .50 \\
            Binary Execution (EM) & 4432 & .40 & .35 & .37 & .37 & .41 & .39 \\
            Encryption-Based Obfuscation (DE) & 4385 & \textbf{.93} & .45 & .60 & \textbf{.91} & .60 & .72 \\
            Suspicious Dependency (MM) & 4385 & .50 & .21 & .30 & .64 & .33 & .44 \\
            Suspicious Domain Exfiltration (EX) & 4208 & .31 & .16 & .21 & .33 & .16 & .22 \\
            Payload Dropper (NO) & 4021 & .16 & .49 & .24 & .16 & .59 & .25 \\
            Description Anomaly (MM) & 3531 & .29 & .59 & .39 & .34 & .68 & .45 \\
            Computational Obfuscation (DE) & 2924 & .41 & .48 & .44 & .32 & .57 & .41 \\
            Import-Time Execution (ES) & 2774 & .10 & .20 & .14 & .11 & .25 & .16 \\
            Binary Dropper (NO) & 2748 & .25 & \textbf{.83} & .38 & .25 & \textbf{.83} & .38 \\
            Combosquatting (MM) & 2733 & .17 & \underline{.03} & \underline{.05} & .20 & \underline{.04} & .07 \\
            Script File Execution (EM) & 1968 & .33 & .51 & .40 & .31 & .59 & .41 \\
            File Deletion (SI) & 1923 & .79 & .40 & .53 & .86 & .52 & .65 \\
            Directory Enumeration (SI) & 1823 & .31 & .54 & .40 & .27 & .56 & .36 \\
            Dynamic Package Install (EM) & 1790 & .38 & .43 & .41 & .40 & .47 & .44 \\
            Reverse Shell (NO) & 1402 & .31 & .70 & .43 & .39 & .71 & .50 \\
            File Relocation (SI) & 1280 & .78 & .35 & .49 & \textbf{.90} & .38 & .54 \\
            Embedded String Payload (DE) & 966 & .12 & .07 & .09 & .10 & .10 & .10 \\
            SSL Validation Bypass (NO) & 913 & .05 & \underline{.02} & \underline{.03} & \underline{.03} & \underline{.01} & \underline{.02} \\
            Sensitive Path Write (SI) & 830 & .18 & .34 & .23 & .18 & .41 & .25 \\
            Environment Modification (SI) & 591 & .17 & .07 & .10 & .38 & .15 & .21 \\
            Archive Dropper (NO) & 571 & .75 & .56 & .64 & .80 & .70 & .75 \\
            File Exfiltration (EX) & 557 & .18 & .38 & .25 & .19 & .48 & .28 \\
            Mining Pool Connection (NO) & 285 & .73 & .68 & \textbf{.70} & \textbf{.88} & .75 & \textbf{.81} \\
            Webhook Exfiltration (EX) & 275 & .19 & .45 & .27 & .24 & .62 & .34 \\
            DNS Tunneling (EX) & 247 & .27 & .24 & .25 & .40 & .21 & .28 \\
            Geolocation Lookup (NO) & 216 & .10 & .20 & .13 & .13 & .29 & .18 \\
            Script Dropper (NO) & 164 & \underline{.04} & .55 & .07 & .05 & .67 & .09 \\
            ASCII Art Deception (DE) & 157 & .43 & .54 & .48 & .53 & .67 & .59 \\
            Startup File Persistence (SI) & 131 & \underline{.04} & .28 & .07 & \underline{.03} & .22 & \underline{.05} \\
            Crypto Wallet Harvesting (SI) & 127 & .14 & .59 & .23 & .16 & .60 & .26 \\
            Unencrypted Communication (NO) & 120 & \underline{.02} & .47 & \underline{.05} & \underline{.02} & .60 & \underline{.05} \\
            \bottomrule
        \end{tabular}
        \\[3pt]
        \raggedright \centering \scriptsize Note: Precision (P), Recall (R), $F_1$ Score ($F_1$). Combination of Precision and Recall that generated highest $F_1$ (Best). \textbf{Bold} = top 3 value in column, \underline{underline} = bottom 3 value in column. Table sorted in descending order by Support.
        \vspace{-5pt}
    \end{table}
    
    Multiple indicators show high precision ($0.85$ to $0.93$) but low recall ($0.19$ to $0.45$).
    For example, \textit{Encryption-Based Obfuscation} shows $0.93$ precision and $0.45$ recall. 
    Recall drops even further to $0.19$ for \textit{Error Suppression}, meaning the model misses 81\% of these instances. 
    This trend demonstrates that \glspl{llm} avoid false positives at the cost of missing true indicators.
    \textit{Binary Dropper} and \textit{Reverse Shell} reverse this trend, prioritizing recall over precision. 
    For \textit{Binary Dropper}, \glspl{llm} produce 75\% false positives but miss only 17\% of actual indicators. 
    Similarly, for \textit{Reverse Shell} (precision 0.31, recall 0.70), models generate 69\% false positives while missing 30\% of indicators.
    The eight worst-performing indicators show mean $F_1 \le 0.10$. 
    Despite a large sample size (5,900 instances), \textit{Dynamic Module Import} shows 0.21 precision and 0.03 recall, missing 97\% of actual indicators. 
    Performance drops further for \textit{SSL Validation Bypass} (precision 0.05, recall 0.02), where \glspl{llm} produce 95\% false positives and miss 98\% of actual indicators.

    \textit{Coder vs General Models: }
    The two code-specialized \glspl{llm} --- \gls{devst} ($F_1=0.75$) and \gls{qwenc} ($F_1=0.66$) --- show a mean weighted $F_1$ score of $0.71$ for category recognition.
    In contrast, the eleven general-purpose \glspl{llm} show a lower mean weighted $F_1$ score of $0.64$.
    The general-purpose \glspl{llm} show high variance, ranging from a weighted $F_1$ score of $0.39$ for \gls{grant} to $0.79$ for \gls{gem31}.
    For indicator identification, the code-specialized \glspl{llm} show a mean weighted $F_1$ score of $0.51$, where \gls{devst} evaluates to a weighted $F_1$ score of $0.56$, and \gls{qwenc} evaluates to $0.46$.
    The general-purpose \glspl{llm} show a lower mean weighted $F_1$ score of $0.41$ compared to code-specialized \glspl{llm}.
    The individual scores of general-purpose \glspl{llm} range from $0.15$ for \gls{grant} to $0.65$ for \gls{gem31}.
    The scores show that specialized code training produces higher $F_1$ score in both category recognition and indicator identification.
    While performance declines across all \glspl{llm} when transitioning from category recognition to indicator identification, code-specialized \glspl{llm} retain a higher mean.

    \begin{takeawaybox}
    \textbf{Key Takeaway:} 
    Under optimal configuration, detection shows mean $F_1=0.91$, compared to $0.75$ for category recognition and $0.52$ for indicator identification, indicating that performance degrades with granularity.
    Detection is recall-biased, showing more false positives.
    Identification is precision-biased, leaving more actual indicators unidentified.
    Prompt strategy improves detection $F_1$ score upto $0.51$ (\gls{fs} $0.23$ to \gls{zs} $0.74$) but its effect on identification is negligible ($\eta^2=0.01$).
    Coder models underperform general-purpose models in detection ($F_1=0.51$ vs. $0.59$) but outperform them in identification ($0.51$ vs. $0.41$).
    \end{takeawaybox}

    \subsection{Findings of $RQ_2$}

    \noindent To address $RQ_2$, we classify all \gls{llm} indicator identification responses into a taxonomy based on the presence or absence of \glspl{tp}, \glspl{fp}, and \glspl{fn} (Table~\ref{tab:outcome_counts}).

    \begin{table}[htbp]
        \centering
        \caption{Indicator Identification Outcomes}
        \vspace{-5pt}
        \label{tab:outcome_counts}
        \setlength{\tabcolsep}{12pt}
        \begin{tabular}{lrr}
            \toprule
            \textbf{Outcome} & \textbf{Count} & \textbf{\%} \\
            \midrule
            CORRECT\_CLEAN & 0 & 0.0\% \\
            CORRECT\_HIT & 135 & 0.3\% \\
            SILENT\_MISS & 367 & 0.7\% \\
            FALSE\_ALARM & 0 & 0.0\% \\
            UNDER\_DETECTION & 8,256 & 16.2\% \\
            OVER\_DETECTION & 1,092 & 2.1\% \\
            MIXED & 37,600 & 73.9\% \\
            MISMATCH & 3,442 & 6.8\% \\
            \midrule
            Total Error (Excluding CORRECT\_HIT) & 50,757 & 99.7\% \\
            \bottomrule
        \end{tabular}
        \vspace{-5pt}
    \end{table}

    \textit{Partial Correctness Over Zero Correct Predictions: }
    The ground-truth set contains the actual indicators present in a package, while the prediction set contains the indicators the \gls{llm} identifies.
    Exact-set matching requires the prediction set to match the ground-truth set completely.
    As ground-truth sets often contain eight to nine simultaneous indicators, exact matches occur infrequently.
    \glspl{llm} identify zero correct indicators in $7.5\%$ of responses (\textit{SILENT\_MISS} $0.7\%$, \textit{MISMATCH} $6.8\%$).
    On the contrary, \glspl{llm} identify at least one correct indicator in $92.2\%$ of responses (\textit{UNDER\_DETECTION} $16.2\%$, \textit{OVER\_DETECTION} $2.1\%$, \textit{MIXED} $73.9\%$).
    The scores indicate that \glspl{llm} mostly predicts one or more correct indicators, and rarely produces zero correct predictions.

    \textit{Missed Indicators: }
    Missed indicators are actual indicators present in a package that the \glspl{llm} leave unidentified.
    To analyze these missed indicators, we divide indicators into two categories: recognizable syntax indicators and contextual indicators.
    Recognizable syntax indicators use standard coding functions, such as executing scripts or modifying files.
    The code itself provides the evidence, and \gls{llm} only needs to identify specific code patterns.
    Contextual indicators, however, require external knowledge or guessing the author's true intent.

    \begin{table}[htbp]
        \vspace{-5pt}
        \centering
        \caption{Top 10 Most-Missed Indicator Techniques}
        \vspace{-5pt}
        \label{tab:most_missed_techniques}
        \begin{tabular}{llc}
            \toprule
            \textbf{Most-Missed Technique} & \textbf{Tactic} & \textbf{Miss Rate} \\
            \midrule
            Combosquatting & Metadata Manipulation & 97.2\% \\
            Dynamic Module Import & Execution Mechanism & 97.0\% \\
            Decoy Functionality & Metadata Manipulation & 87.9\% \\
            Suspicious Domain Exfiltration & Exfiltration & 84.5\% \\
            Error Suppression & Defense Evasion & 81.3\% \\
            Suspicious Dependency & Metadata Manipulation & 78.7\% \\
            Conditional Payload Trigger & Execution Mechanism & 77.2\% \\
            Typosquatting & Metadata Manipulation & 74.9\% \\
            Encoding-Based Obfuscation & Defense Evasion & 69.8\% \\
            Suspicious Author Identity & Metadata Manipulation & 60.3\% \\
            \bottomrule
        \end{tabular}
        \vspace{-5pt}
    \end{table}

    Mapping the most-missed indicators (see Table~\ref{tab:most_missed_techniques}) to their corresponding categories shows that most misses cluster within \textit{Metadata Manipulation} and \textit{Defense Evasion}.
    These categories consist entirely of contextual indicators that require information unavailable to \glspl{llm}.
    Identifying the contextual indicators requires information that is not available to \glspl{llm}.
    For example, identifying \textit{Typosquatting} and \textit{Combosquatting} requires an external popularity index (such as download statistics) to recognize when a malicious package name intentionally mimics a popular legitimate package.
    Similarly, identifying a \textit{Suspicious Author} requires external threat intelligence to verify if an author account has known malicious history.
    
    Other contextual indicators require subjective analysis of the author's goals.
    Identifying \textit{Decoy Functionality} requires such analysis. 
    If the code executes as documented, \glspl{llm} cannot distinguish a legitimate feature from a payload.
    Furthermore, identifying \textit{Error Suppression} and obfuscation related indicators require contextual judgment to distinguish between poor developer practices, such as broad exception handling, and an attacker deliberately silencing failed malicious executions.
    
    Conversely, miss proportions decrease for recognizable syntax indicators, such as \textit{Install-Time Execution} (miss rate $47.8\%$), \textit{Arbitrary File Write} ($43.7\%$), \textit{Hidden Code Execution} ($34.1\%$), and \textit{Shell Command Execution} ($29.3\%$).
    Additionally, the near-total miss proportions for \textit{Dynamic Module Import} ($97.0\%$) and \textit{Combosquatting} ($97.2\%$) indicate that these contextual indicators are rarely identified by \glspl{llm}.

    \textit{Hallucinated Indicators: }
    Hallucination occurs when a \gls{llm} incorrectly identifies an indicator that does not actually exist in the code, generating a false positive.
    Mapping the most-hallucinated indicators (Table~\ref{tab:most_hallucinated_techniques}) show a clustering in specific categories, similar to missed indicators. 
    These false positives occur primarily within \textit{Network Operations} (e.g., payload droppers, external connections) and the \textit{Execution Stage} (e.g., \textit{Install-Time Execution} and \textit{Import-Time Execution}).

    \begin{table}[htbp]
        \vspace{-5pt}
        \centering
        \caption{Most-Hallucinated Indicators}
        \vspace{-5pt}
        \label{tab:most_hallucinated_techniques}
        \begin{tabular}{llr}
            \toprule
            \textbf{Most-Hallucinated Technique} & \textbf{Tactic} & \textbf{FP Count} \\
            \midrule
            Install-Time Execution & Execution Stage & 12,109 \\
            Payload Dropper & Network Operations & 10,554 \\
            Binary Dropper & Network Operations & 7,004 \\
            Suspicious Connection & Network Operations & 5,700 \\
            Arbitrary File Write & System Impact & 5,339 \\
            Description Anomaly & Metadata Manipulation & 5,099 \\
            Import-Time Execution & Execution Stage & 5,047 \\
            Dynamic Evaluation & Execution Mechanism & 3,914 \\
            \bottomrule
        \end{tabular}
        \vspace{-5pt}
    \end{table}

    Hallucinations occur because \glspl{llm} scan keywords instead of analyzing code. 
    When \glspl{llm} encounter common programming functions—such as network requests (\texttt{requests.get}), installation files (\texttt{setup.py}), or file writing commands (\texttt{open().write()})—they assume that the code is malicious. 
    As attackers frequently use these recognizable syntax, the \glspl{llm} learn to identify those during training. 
    \glspl{llm} make this decision based on the presence of the syntax, ignoring the surrounding context and the actual goal of the program.
    For example, if a \gls{llm} encounters a script downloading a safe program, the \gls{llm} often misclassifies the action as a \textit{Payload Dropper} because it detects a network command.
    Similarly, identifying standard setup functions inside a \texttt{setup.py} file often causes \glspl{llm} to report an \textit{Install-Time Execution} attack. 
    This behavior shows that \glspl{llm} fail to trace data flow or understand program intent, relying entirely on matching basic programming patterns to previous examples of malware.
    
    \textit{Reasons behind \textit{MISMATCH}: }
    A \textit{mismatch} occurs when the prediction and ground-truth share zero indicators. 
    Analyzing the responses of 13 \glspl{llm}, we identify that \textit{mismatch} outcomes stem from three reasons: ground-truth error, granularity mismatch, and genuine misses. 
    Some representative examples from \gls{qwen4} at \gls{zs} and T=0.0 are provided below:

\begin{itemize}[leftmargin=10pt]
    \item Ground-Truth Error (Measurement Limitation): The dataset is incorrect or incomplete.
    \begin{itemize}[leftmargin=5pt]
        \item \texttt{secretinspector-0.1.0}: 
        \gls{gt} = \{\textit{Lifecycle Hook Hijack}, \textit{Reverse Shell}\}; Prediction = \{\textit{Shell Command Execution}\}. 
        The \gls{gt} flags a commented-out reverse shell, while the \gls{llm} identifies a live shell command (\texttt{os.system("echo 'You Have been pwned' > /tmp/pwned")}). 
        The \gls{llm} prediction is more accurate than the ground-truth.
    \end{itemize}
    
    \item Granularity Mismatch (Capability Limitation): 
    The \gls{llm} identifies the correct indicator category but selects another indicator within the same category.
    \begin{itemize}[leftmargin=5pt]
        \item \texttt{upggrade-requests-0.0.1}: 
        \gls{gt} = \{..., \textit{Combosquatting}\}; Prediction = \{\textit{Typosquatting}\}. 
        \gls{llm} recognizes category but selects an adjacent indicator.
        \item \texttt{cipherbcrypt-1.4}: \gls{gt} = \{\textit{Computational Obfuscation}, \textit{Data Exfiltration}\}; 
        Prediction = \{\textit{Encoding-Based Obfuscation}, \textit{Suspicious Domain Exfiltration}, \textit{Payload Dropper}\}. 
        \gls{llm} recognizes obfuscation and exfiltration categories but selects adjacent indicators.
    \end{itemize}
    
    \item Genuine Misses (Capability Limitation): 
    The \gls{llm} leave the actual indicator unidentified.
    \begin{itemize}[leftmargin=5pt]
        \item \texttt{j3y5r-0.0.0}: \gls{gt} = \{\textit{Computational Obfuscation}, \textit{Import-Time Execution}, \textit{Dynamic Evaluation}, \textit{Suspicious Author Identity}\}; 
        Prediction = \{\textit{Typosquatting}\}. 
        Execution and obfuscation behaviors stay undetected.
    \end{itemize}
\end{itemize}

    \textit{Prompt Strategy Changes Outcome Distribution: }
    A perfect prediction occurs when \gls{llm} identifies the exact set of actual indicators in a package. 
    Changing the prompt strategy does not increase perfect predictions, $99.7\%$ of all outcomes remain imperfect. 
    However, different prompts change the distribution of these imperfect outcomes (specifically UNDER\_DETECTION, MIXED, and MISMATCH) by shifting \glspl{llm}' willingness to guess versus remain silent.

    \begin{table}[htbp]
        \vspace{-5pt}
        \centering
        \caption{Outcome Distribution by Prompt Strategy}
        \vspace{-5pt}
        \label{tab:prompt_outcomes}
        \setlength{\tabcolsep}{3.5pt}
        \begin{tabular}{lrrr}
            \toprule
            \textbf{Prompt Strategy} & \textbf{\texttt{UNDER\_DETECTION}} & \textbf{\texttt{MIXED}} & \textbf{\texttt{MISMATCH}} \\
            \midrule
            \texttt{zero\_shot} & 2,938 (highest) & 7,661 & 811 \\
            \texttt{chain\_of\_thought} & 1,756 & 7,855 & 626 \\
            \texttt{few\_shot} & 2,110 & 8,213 & 1,121 (highest) \\
            \texttt{tree\_of\_thought} & 905 & 5,877 & 474 \\
            \texttt{self\_consistency} & 547 (lowest) & 7,994 & 410 \\
            \bottomrule
        \end{tabular}
        \vspace{-5pt}
    \end{table}

    The \gls{zs} prompt makes the model cautious. 
    It produces the highest rate of \textit{UNDER\_DETECTION} outcomes because 
    \gls{llm} avoids predicting when uncertain, which leaves actual indicators unidentified.
    Conversely, the \gls{sc} prompt triggers over-predicting. 
    It produces the lowest \textit{UNDER\_DETECTION} outcomes but pushes most outcomes into the \textit{MIXED} category.
    In this state, the \gls{llm} generates false positives (hallucinations) and false negatives (missed indicators) alongside its correct predictions.
    The \gls{fs} prompt shows the highest rate of \textit{MISMATCH} outcomes, producing completely incorrect predictions.

    \begin{table}[htbp]
        \vspace{-5pt}
        \centering
        \caption{Classification based on Investigation Effort}
        \label{tab:investigation-effort}
        \begin{tabular}{llc}
            \toprule
            \textbf{Effort} & \textbf{Associated Outcome} & \textbf{Total (\%)} \\
            \midrule
            Low & OVER\_DETECTION & 2.1 \\
            Medium & UNDER\_DETECTION & 16.2 \\
            High & MIXED & 73.9 \\
            Critical & MISMATCH (6.8\%), \textit{SILENT\_MISS} (0.7\%) & 7.5 \\
            \bottomrule
        \end{tabular}
        \vspace{-5pt}
    \end{table}

    \textit{Classification based on Investigation Effort: }
    We categorize the outcomes based on subsequent investigation effort required by human analyssts after using \glspl{llm} for initial triage (Table~\ref{tab:investigation-effort}.
    \textit{OVER\_DETECTION} (2.1\%) captures all actual indicators but introduces false positives, which requires low effort, as analysts must verify and discard invalid indicators.
    \textit{UNDER\_DETECTION} (16.2\%) misses some actual indicators but avoids false positives, which requires medium effort, as analysts must search for the remaining indicators.
    \textit{MIXED} (73.9\%) misses actual indicators and introduces false positives, requires high effort, as analysts must discard invalid indicators and simultaneously search for remaining indicators.
    \textit{MISMATCH} (6.8\%) and \textit{SILENT\_MISS} (0.7\%) provide zero actual indicators and may add false positives, which requires critical effort, forcing analysts to discard invalid indicators and conduct a full investigation from scratch.
    Consequently, with over 80\% of outcomes requiring high or critical human effort, \glspl{llm} do not add much value as triage filters for indicator identification.

    \begin{takeawaybox}
    \textbf{Key Takeaway:} 
    Responses predominantly show partial correctness: $92.2\%$ contain at least one correct indicator. 
    $73.9\%$ of all responses show \textit{MIXED} outcome.
    False negatives predominantly occur in contextual categories that require external knowledge, which \glspl{llm} do not have.
    False positives predominantly occur in categories reliant on recognizable syntax, where \glspl{llm} predict common code patterns used by adversaries as malicious ignoring surrounding context.
    Prompt strategy changes the outcome distribution: \gls{zs} produces conservative prediction, \gls{sc} triggers over-prediction, and \gls{fs} generates the highest proportion of incorrect predictions.
    \end{takeawaybox}

    \subsection{Findings of $RQ_3$}
    
    When evaluating data, a measurable difference is not always a meaningful one. 
    Statistical significance, measured by a $p$-value, indicates whether an observation is reliable or just random noise. 
    $p$-value less than $0.05$ means that the result is due to a genuine correlation rather than random chance.
    However, a correlation can be statistically significant without being practically useful.
    To determine if the finding actually matters, we measure the strength of that correlation.

    To measure the strength of the correlation, we use statistical tools.
    We use Spearman's correlation ($\rho$) to evaluate the relationship between two variables on a scale from $-1$ to $1$. 
    A positive value indicates that both variables increase together, a negative value indicates that one increases while the other decreases, and a value near zero indicates no meaningful relationship. 
    For categorical variables, we use Eta-squared ($\eta^2$). 
    This metric ranges from $0$ to $1$ and measures the percentage of the final outcome explained by that variable.
    
    In Table~\ref{tab:rq3-summary}, we measure the correlation between indicator-identification $F_1$ score and five factors: parameter size, context width, prompt strategy, temperature, and code complexity.

    \textit{Parameter Size: }
    Across the 12 \glspl{llm}, increasing the parameter size shows a positive correlation with the $F_1$ score ($\rho=+0.431$). 
    However, this correlation lacks statistical significance ($p=0.16$), meaning the observation could simply be random noise rather than a genuine correlation.
    Individual \gls{llm} results demonstrate this lack of statistical significance. 
    Two \glspl{llm} with roughly the same size (32 to 33 billion parameters) achieved the highest (0.65) and lowest (0.17) scores. 
    Furthermore, a smaller 4-billion parameter \gls{llm} (0.42) outperformed a larger 15-billion parameter \gls{llm} (0.29).
    As the test included 12 \glspl{llm}, the sample size is too small to confirm a reliable correlation. 
    Hence, there is no evidence that a larger model size guarantees better performance.

    \begin{table}[htbp]
        \caption{Factors Influencing Indicator-Identification $F_1$}
        \label{tab:rq3-summary}
        \centering
        \begin{tabular}{lccl}
            \hline
            Factor & Statistic & $p$ & Effect size \\
            \hline
            Parameter size     & $\rho=+0.43$    & $0.16$     & Moderate (n.s.) \\
            Context width      & $\rho=+0.25$    & $0.41$     & Small (n.s.) \\
            Temperature        & $\rho=-0.06$    & $<0.0001$  & Negligible \\
            Prompt strategy & $\eta^2=0.01$   & $<0.0001$  & Negligible \\
            Code complexity    & $\rho=-0.26$    & $<0.0001$  & Small \\
            \hline
        \end{tabular}
        \\[2pt]
        {\footnotesize n.s.\ = not significant. Prompt strategy uses Kruskal--Wallis
        ($H=439.18$); the other factors use Spearman $\rho$.}
    \end{table}

    \textit{Context Width: }
    Across the 13 \glspl{llm} evaluated, context width shows a weak positive correlation ($\rho=+0.25$) with the $F_1$ score. 
    However, this correlation is not statistically significant ($p=0.41$), suggesting the observation could be random noise rather than a genuine correlation.
    The analysis is limited by a lack of variation in the data, as most \glspl{llm} use a context window of at least $109K$ tokens. 
    The positive correlation is largely driven by one \gls{llm} with a smaller window ($16K$), which also recorded one of the lowest scores.
    As most \glspl{llm} cluster around similar capacities, there is no dependable evidence in this study that a wider context window improves identification performance.

    \textit{Temperature: }
    Temperature shows a statistically significant correlation with $F_1$ scores, meaning the relationship is genuine rather than random noise ($p < 0.0001$). 
    However, the correlation strength is negligible ($\rho = -0.056$), indicating the correlation lacks practical importance.
    The data shows mean $F_1$ scores decrease slightly as temperature rises, shifting from $0.470$ at $T=0.0$ to $0.449$ at $T=0.5$ and $0.442$ at $T=1.0$. 
    While lower temperatures lead to more consistent labeling, the shift is too small to matter in practice. 
    Therefore, temperature does not meaningfully impact identification performance.

    \textit{Prompt Strategy: }
    Prompt strategy has a statistically significant correlation with $F_1$ scores, indicating a genuine relationship rather than random noise ($p < 0.0001$). 
    However, the effect size is negligible ($\eta^2 = 0.0086$, which falls below the $0.01$ threshold for a small effect), meaning the choice of prompt strategy lacks practical importance.
    Self-consistency (\gls{sc}) attains the highest mean $F_1$ ($0.486$), while zero-shot (\gls{zs}) attains the lowest ($0.431$), representing a difference of approximately $0.055$. 
    Other strategies, such as Chain-of-Thought (\gls{cot}, $0.463$) and Tree-of-Thoughts (\gls{tot}, $0.462$), do not exceed the performance of self-consistency. 
    These results are specific to the indicator-identification task, the more substantial impact of prompt strategy observed in binary package detection (Section~\ref{section:findings-rq1}) does not apply here.

    \textit{Code Complexity: }
    Code complexity has the strongest correlation with $F_1$ scores, indicating an actual relationship rather than random noise ($p < 0.0001$). 
    Furthermore, the correlation strength is small but meaningful ($\rho = -0.26$), confirming that this factor has the most significant impact on model performance among all factors.
    Across the $370$ packages, a higher \gls{loc} is consistently associated with lower $F_1$ scores. 
    This decline is observed among the largest packages: small ($\leq 33$ \gls{loc}, mean $F_1$ $0.47$) and medium ($33$--$34$ \gls{loc}, $0.49$) packages perform similarly, whereas large packages ($>34$ \gls{loc}, $0.38$) show a larger drop. 
    Larger packages contain more indicators and offer more opportunities for performance drop.

    \begin{takeawaybox}
    \textbf{Key Takeaway:} 
    No \gls{llm} setting (parameter size, context width, temperature, or prompt strategy) has a practical effect on indicator-identification: performance cannot be improved by adjusting \gls{llm} settings.
    Only code complexity has significant correlation with indicator identification: performance degrades in larger packages.
    \end{takeawaybox}

    \section{Discussion}

    In this section, we discuss the implications for practitioners and researchers.

    \subsection{Implications for Practitioners}
    
    \glspl{llm} should be used as an initial triage tool to filter malicious packages because models rarely miss a malicious packages (only $1.6\%$ of configurations show $<0.5$ recall).
    High recall makes \glspl{llm} well-suited to flag suspicious packages before human review.
    However, \glspl{llm} frequently flag safe packages as malicious ($51\%$ of configurations show $<0.5$ precision), so a human must confirm every flagged package before any action is taken.
    \glspl{llm} should not be trusted to explain why a package is malicious.
    At the indicator level, $73.9\%$ of responses are \textit{MIXED}, meaning they contain both correct indicators and false positives, and $>80\%$ of responses require high or critical analyst effort to clean up.
    If an analyst trusts an \gls{llm} blindly, he will chase fake alerts and miss actual indicators of malicious package.

    \glspl{llm} should be paired with standard tools to cover blind spots.
    \glspl{llm} miss contextual threats because they lack external data. 
    Instead, analysts should use basic checks: a popularity index for \textit{Typosquatting}, a reputation lookup for \textit{Suspicious Author}, and dataflow analysis to verify if a network call is actually malicious.
    Larger models, wider context windows, and complex prompts do not improve performance. 
    Therefore, practitioners should deploy small, cost-effective \glspl{llm} alongside standard tools rather than relying on massive models. 
    Additionally, standalone $F_1$ scores are uninformative because performance can fluctuate by $\Delta F_1 = 1.0$. 
    Finally, any reported $F_1$ score is meaningless without its configuration, since the same \gls{llm} can vary by up to $F_1 = 1.0$ across configurations.
    Therefore, practitioners should always record the prompt strategy and temperature used.

    \subsection{Implications for Researchers}

    \gls{llm} performance declines as task becomes more fine-grained: mean $F_1=0.91$ for detection under optimal configuration, $0.75$ for category recognition, and $0.52$ for indicator identification.
    Detection needs only one indicator to return a correct verdict, recognition needs the category of indicator, identification needs specific indicators, and many indicators need external knowledge such as package popularity, author history, or the author's intent.
    As granularity gets finer, the task demands more information, and eventually performance drops.
    This is a limitation of what \glspl{llm} can do, so it motivates supplying the missing information through retrieval over registry, popularity, and authorship data.

    Researchers should evaluate identification performance with metrics that reward partial correctness, not exact-set matching.
    Exact-set matching scores responses with one missed indicator the same as a responses with zero correct indicators, which is misleading because 92.2\% of responses contain at least one correct indicator.
    To evaluate \glspl{llm}, researchers should use overlap-based metrics that gives credit to prediction with one or more correct indicators, and category-aware metrics that gives credit to prediction with wrong indicator but right category.

    \section{Threats to Validity}

    \textit{Construct Validity: }
    Exact-set match penalizes partial correctness, so we also report the full outcome taxonomy and indicator scores.
    Incomplete or subjective ground-truth labels may artificially lower performance scores.
    \textit{Internal Validity: }
    To control for \gls{llm} randomness and prompt sensitivity, we used fixed prompt templates.
    As the test packages are public, models may have seen them during training. 
    While this memorization might artificially inflate detection scores, the models' failure to identify specific indicators proves it does not grant actual code understanding.
    \textit{External Validity: }
    Our study targets \gls{pypi} and Python and may not transfer to ecosystems such as \gls{npm} or RubyGems. 
    The second task uses only malicious packages, and the benign corpus is restricted to small packages to fit context windows.
    \textit{Conclusion Validity: }
    Our small sample size of 12 to 13 models limits statistical power. 
    Consequently, we prioritize effect sizes over $p$-values and make no causal claims regarding model specifications.

    \section{Conclusion}

    We evaluated 13 \glspl{llm} for detecting malicious \gls{pypi} packages and identifying their fine-grained indicators across a 47-indicator taxonomy, five prompt strategies, and three temperatures. 
    We observe that the performance of \glspl{llm} degrades with granularity.
    \glspl{llm} predominantly produce partially correct responses.
    \glspl{llm} miss context dependent indicators and generates false positives for indicators with recognizable syntax.
    Prompt strategy changes the outcome distribution.
    Finally, performance degrades with increasing code complexity.
    These findings motivate overlap-based and hierarchical evaluation metrics, \gls{llm} triage paired with standard tools and information retrieval, fine-tuning and agentic designs to improve indicator identification performance, and extension to other ecosystems.

    \section*{Declaration of Generative AI Technologies }
    During manuscript preparation, the authors used ChatGPT only to improve flow, grammar, and clarity. The tool was not used to generate technical content, synthesize citations, or verify experimental facts. The authors reviewed and verified all outputs and take full responsibility for the final manuscript.

    \bibliographystyle{IEEEtran}
    \bibliography{references}

\end{document}